\documentclass[prl,twocolumn,floatfix,nofootinbib,superscriptaddress,
longbibliography]{revtex4-1}

\usepackage{graphicx}
\usepackage[fleqn]{amsmath}
\usepackage{amssymb}
\usepackage[usenames]{color}
  \definecolor{goodgreen}{rgb}{0.1,0.5,0}
  \definecolor{goodred}{rgb}{0.7,0,0}
\usepackage{pifont}
\usepackage{float}
\usepackage{multirow}
\usepackage{booktabs}
\usepackage[colorlinks,urlcolor=blue,citecolor=goodgreen,linkcolor=goodred]%
{hyperref}
\usepackage{bm}
\usepackage[normalem]{ulem}

\usepackage{mathptmx}
\usepackage[scaled=.90]{helvet}
\usepackage{courier}

\allowdisplaybreaks
\newcommand{\ve}[1]{\ensuremath{\bm{#1}}}

\newcommand{\vg}{\ensuremath{V_\text{gate}}}
\newcommand{\vb}{\ensuremath{V_\text{bias}}}
\newcommand{\didv}{\ensuremath{\text{d}I / \text{d}\vb}}
\newcommand{\un}[1]{\ensuremath{\,\text{#1}}}
\newcommand{\Bpar}{\ensuremath{B_\parallel}}

\newcommand{\kpar}{\ensuremath{k_\parallel}}
\newcommand{\kperp}{\ensuremath{k_\perp}}

\begin{document}

\title{Shaping electron wave functions in a carbon nanotube with a 
parallel magnetic field}

\author{M. Marga\'{n}ska}
\affiliation{Institute for Theoretical Physics, University of Regensburg, 93053 
Regensburg, Germany} 
\author{D. R. Schmid}
\affiliation{Institute for Experimental and Applied Physics, University of 
Regensburg, 93053 Regensburg, Germany}
\author{A. Dirnaichner}
\affiliation{Institute for Experimental and Applied Physics, University of 
Regensburg, 93053 Regensburg, Germany}
\author{P. L. Stiller}
\affiliation{Institute for Experimental and Applied Physics, University of 
Regensburg, 93053 Regensburg, Germany}
\author{Ch. Strunk}
\affiliation{Institute for Experimental and Applied Physics, University of 
Regensburg, 93053 Regensburg, Germany}
\author{M. Grifoni}
\affiliation{Institute for Theoretical Physics, University of Regensburg, 93053 
Regensburg, Germany} 
\author{A. K. H\"{u}ttel}
\email{andreas.huettel@ur.de}
\affiliation{Institute for Experimental and Applied Physics, University of 
Regensburg, 93053 Regensburg, Germany}

\date{February 23, 2019}

\begin{abstract}
A magnetic field, through its vector potential, usually causes measurable 
changes in the electron wave function only in the direction transverse to the 
field. Here we demonstrate experimentally and theoretically that in carbon 
nanotube quantum dots, combining cylindrical topology and bipartite hexagonal 
lattice, a magnetic field along the nanotube axis impacts also the {\em 
longitudinal} profile of the electronic states. With the high (up to 17\,T) 
magnetic fields in our experiment the wave functions can be tuned all the way 
from ``half-wave resonator'' shape, with nodes at both ends, to ``quarter-wave 
resonator'' shape, with an antinode at one end. This in turn causes a distinct 
dependence of the conductance on the magnetic field. Our results demonstrate a 
new strategy for the control of wave functions using magnetic fields in quantum 
systems with nontrivial lattice and topology.
\end{abstract}

\maketitle

\begin{figure}[tbp]
\includegraphics[width=0.9\columnwidth]{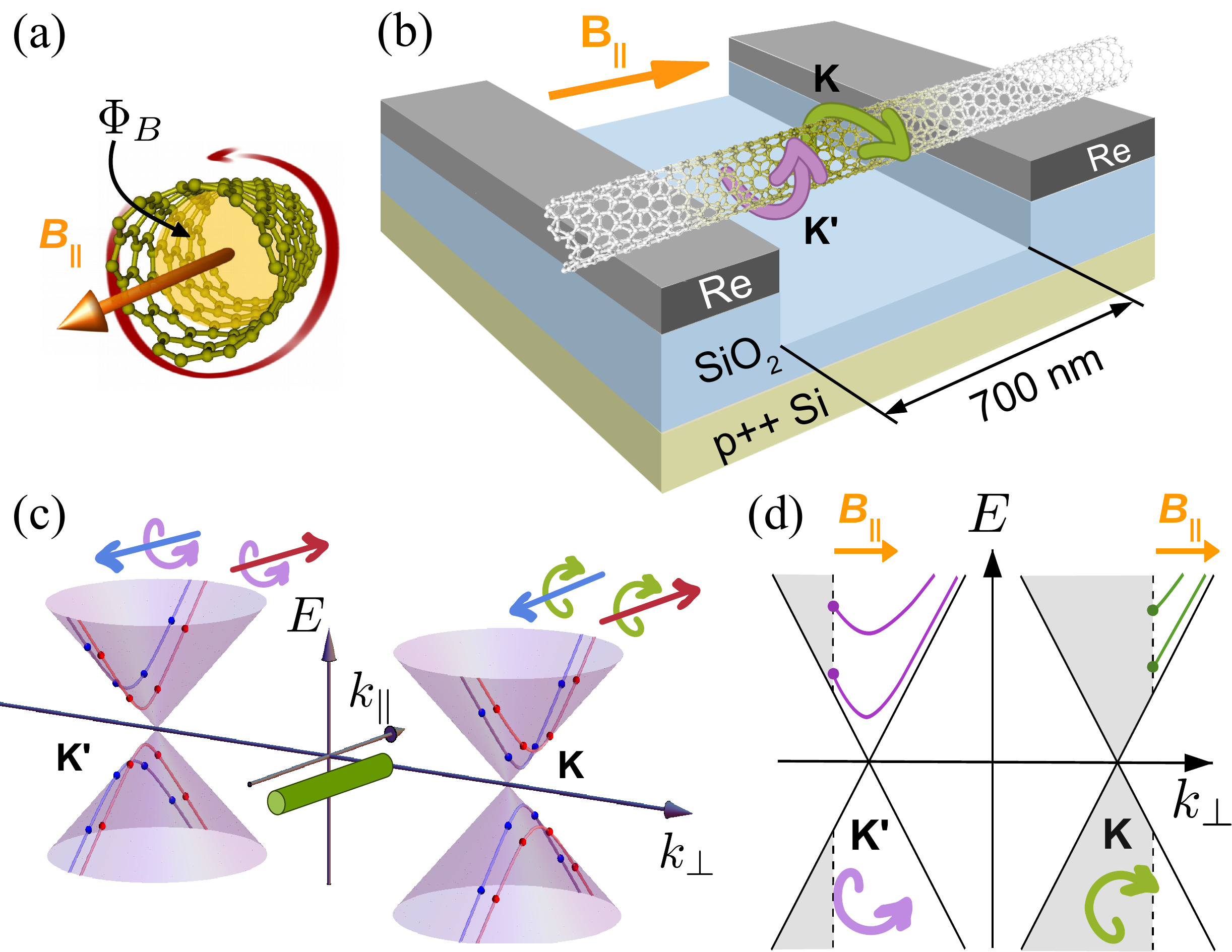} 
\caption{\label{fig:device}
(a) Electrons circulating in closed orbits acquire an Aharonov-Bohm phase 
proportional to the enclosed magnetic flux.
(b) Schematic of a suspended CNT device with its embedded quantum dot (shaded 
green) and a magnetic field parallel to the nanotube. 
(c) Dirac cones of the graphene dispersion relation. Blue and red lines 
indicate the lowermost transverse subbands forming in a CNT. Spin degeneracy is 
lifted by the spin-orbit coupling. Quantized \kpar\ values due to a finite CNT 
length are marked with dots; $\Bpar=0$.
(d) An axial magnetic field changes \kperp\ via the Aharonov-Bohm effect, 
shifting the 1-d subbands across the Dirac cones. 
}
\end{figure}

As first noticed by Aharonov and Bohm \cite{aharonov:pr1959}, when a charged 
quantum particle travels in a finite electromagnetic potential, its wave
function acquires a phase whose magnitude depends on the travelled path. For
particles with electric charge $q$ moving along a closed path, the phase shift
$\varphi_{AB} = q\Phi_B/h$, known as Aharonov-Bohm shift, is expressed in terms
of the magnetic flux $\Phi_B$ across the enclosed area. Because $\Phi_B$
depends only on the magnitude of the magnetic field component normal to this
area's surface, the phase is acquired along directions {\it transverse} to the
magnetic field, see Fig.~\ref{fig:device}(a). In mesoscopic rings or tubular
structures pierced by a magnetic field, the phase changes the quantization
condition for the tangential part of the electronic wave vector by $k_\perp
\rightarrow k_\perp + \varphi_{AB}/r$ (with $r$ the radius of the ring or
tubulus) and is at the basis of remarkable quantum interference phenomena
\cite{prl-webb:2696}. However, as the perpendicular components of the magnetic
vector potential commute with the parallel component of the momentum, a
parallel magnetic field is not expected to affect the wave function along the
field.

Also in carbon nanotubes (CNTs), the electronic wave function acquires an 
Aharonov-Bohm phase when a magnetic field is applied along the nanotube axis  
\cite{ajiki:jpsj1993}, see Fig.~\ref{fig:device}(a). The phase gives rise to  
resistance oscillations in a varying magnetic flux \cite{bachtold:nature1999}. 
Since it changes \kperp, it also changes the energy $E({\bm k})$ of an 
electronic state, through its dependence on the wave vector $\ve{k} = 
(\kpar,\kperp (\Bpar))$. Such a magnetic field dependence of the energies has 
been observed through beatings in Fabry-Perot patterns \cite{cao:prl2004}, or in 
the characteristic evolution of excitation spectra of CNT quantum dots in the 
sequential tunneling \cite{minot:nature2004, kuemmeth:nature2008, 
jespersen:natphys2011, steele:natcomms2013} and Kondo 
\cite{jarillo-herrero:nature2005, paaske:natphys2006, makarovski:prb2007, 
grove-rasmussen:prl2012, schmid:prb2015, niklas:natcomms2016} regimes. 

In this Letter we show that the combination of the bipartite honeycomb lattice, 
the cylindrical topology of the nanotubes, and the confinement in the quantum 
dot intertwines the usually separable parallel and transverse components of the 
wave function. This leads to unusual tunability of the wave function in the
direction {\em parallel} to the magnetic field. Experimentally, it manifests in 
a pronounced variation of the conductance with magnetic field, arising from the 
changes of the wave function amplitude near the tunnel contacts between the
electrostatically defined quantum dot and the rest of the CNT.

Similar to graphene, in CNTs the honeycomb lattice gives rise to two 
non-equivalent Dirac points $\ve{K}$ and $\ve{K'}$ (also known as valleys). The 
valley and spin degrees of freedom characterize the four lowermost CNT 
subbands, see Fig.~\ref{fig:device}(c). Our measurements display i) a 
conductance rapidly vanishing in a magnetic field for transitions associated to 
the $K$-valley; ii) an increase and then a decrease of the conductance for 
$K'$-valley transitions as the axial field is varied from $0$ up to $17\un{T}$. 
Similar behavior can be found in results on other CNT quantum dots, see, e.g.,
Figs.~1(c) and S9 of \cite{steele:natcomms2013} or Fig.~2 of
\cite{deshpande:science2009}. To our knowledge, no microscopic model explaining
it has yet been proposed. Our calculation captures this essential difference
between the K and K' valley states.

{\em{Dispersion relation of long CNTs---}}
In CNTs the eigenstates are spinors in the bipartite honeycomb lattice space, 
solving the Dirac equation, Eq.~(\ref{eq:Hgraphene}) below. The resulting 
dispersion is $E(\ve{k})=\pm \hbar v_F\sqrt{\kappa^2_\parallel + 
\kappa^2_\perp}$, see Fig.~\ref{fig:device}(c), where the 
$\kappa_{\perp/\parallel} = k_{\perp/\parallel} - \tau K_{\perp/ \parallel}$ 
are wave vectors relative to the graphene Dirac points $\ve K$ ($\tau=1$) and 
$\ve{K}'=-\ve{K}$ ($\tau=-1$). 

The cylindrical geometry restricts the values of the transverse 
momentum \kperp\ through the boundary condition $\Psi(\ve{R} +\ve{C}_{}) = 
\Psi(\ve{R})$, with $\ve{C}_{}$ the wrapping vector of the CNT, generating
transverse subbands. Furthermore, curvature causes a chirality-dependent offset 
$\tau\Delta\ve{k}^c$ of the Dirac points, opening a small gap in nominally 
metallic CNTs with $\kappa_\perp=0$, as well as a spin-orbit coupling induced 
shift $\sigma k_{SO}$ of the transverse momentum \cite{izumida:jpsj2009, 
klinovaja:prb2011, laird:rmp2015} ($\sigma=\pm 1$ denotes the projection of the 
spin along the CNT axis). As shown in Fig.~\ref{fig:device}(c), the latter 
removes spin-degeneracy of the transverse subbands. When an axial magnetic field 
is applied, the Aharonov-Bohm phase further modifies \kperp. The energy 
$E(\kpar, \kperp(\Bpar))$ of an infinite CNT then follows again from the Dirac 
equation under the replacements
\begin{align} \nonumber
\kperp &\to \kperp + \frac{\varphi_{AB}}{r} +\sigma \Delta 
k_{SO}+\tau 
\Delta k^c_\perp,\\ \label{kappapar}
\kpar &\to \kpar + \tau \Delta 
k^c_{\parallel}, 
\end{align}
the addition of a Zeeman term $\mu_\text{B} \sigma \Bpar$, and a 
field-independent energy shift due to the spin-orbit coupling 
\cite{izumida:jpsj2009, klinovaja:prb2011, laird:rmp2015}. In CNT quantum 
dots with lengths of few hundreds of nanometers the longitudinal wave vector 
becomes quantized, leading to discrete bound states (dots in
Fig.~\ref{fig:device}(c)). The magnetic field dependence of $E$ for two bound 
states belonging to different valleys is shown in Fig.~\ref{fig:device}(d) for 
fixed \kpar. A characteristic evolution, distinct for the two valleys, is 
observed. 

{\em{Magnetospectrum of a CNT quantum dot---}}
Fig.~\ref{fig:device}(b) shows a schematic of our device: a suspended CNT grown
{\it in situ} over rhenium leads \cite{nature-kong:878, cao:natmater2005}.
Tuning the back gate voltage we can explore both hole and electron conduction.
As typical for growth over rhenium or platinum electrodes, the metal-CNT
contacts are transparent, and the CNT is effectively p-doped near them. In the
electron conduction regime, gating then causes two p-n junctions within the
CNT, which, as tunnel barriers, lead to Coulomb blockade \cite{apl-park-2001,
minot:nature2004, nnano-steele-2009}. We can clearly identify the gate voltage
region corresponding to $0 \le N \le 1$ trapped conduction band electrons; an
electron is here confined to a fraction of the $700\un{nm}$ metal contact
distance, with the rest of the CNT acting as barriers and leads. From
the spectrum, we estimate a confinement length $L\sim 400\un{nm}$ or $L\sim
240\un{nm}$ depending on the method used (see Sec.\ III of the Supplement
\cite{supplement} for details).

\begin{figure}[tbp]
\includegraphics[width=0.9\columnwidth]{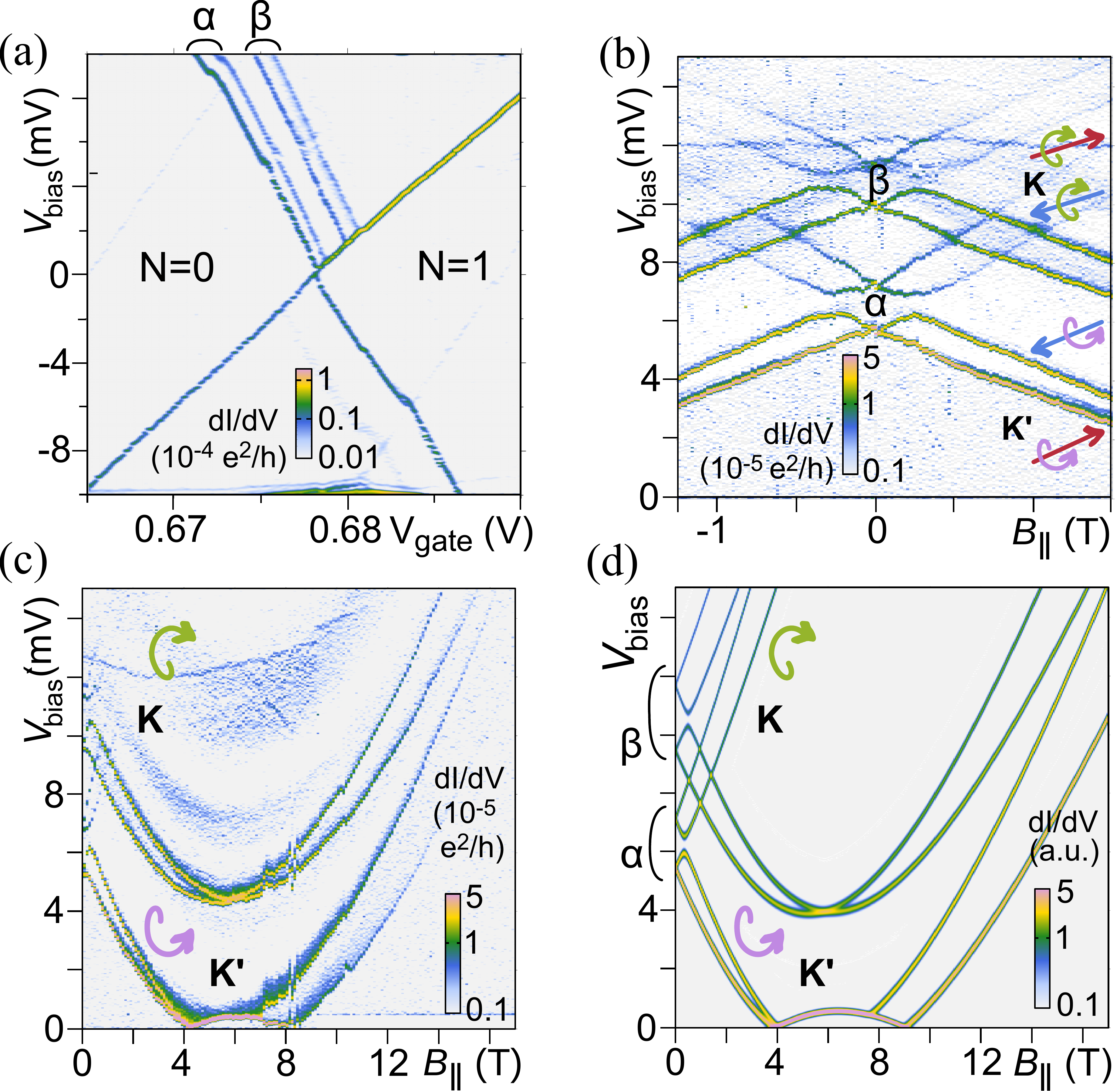} 
\caption{\label{fig:data}
(a) Zero magnetic field differential conductance \didv\ of a CNT quantum dot
with $0\le N \le 1$ conduction band electrons. Two pairs $\alpha$ and $\beta$
of conductance lines, all four representing Kramers doublets, are visible.
(b) \didv\ for constant $\vg = 0.675\un{V}$ and varying $|\Bpar|
\le 1.5\un{T}$. The Kramers doublets split at finite field into four states for
both $\alpha$ and $\beta$. Spin and valley of the $\alpha$ states for $\Bpar
\gg 0.5\un{T}$ are indicated.
(c) Differential conductance at the same \vg, now for \Bpar\ up to $17\un{T}$. 
The four visible lines correspond to $K'$ states in $\alpha$ and
$\beta$; the $K$ lines fade out fast. 
(d) Calculated conductance, using the reduced density matrix technique and 
assuming {\it field-independent} tunneling coupling of all states to the leads. 
In contrast to the measurement, both $K$ and $K'$ valley lines clearly persist 
at high magnetic field.
}
\end{figure}
Figure~\ref{fig:data}(a) shows the stability diagram of the CNT in this gate 
voltage region. The resonance lines correspond to the single particle energies 
of the lowest discrete states of the quantum dot \cite{laird:rmp2015}. Two
closely spaced sets $\alpha$ and $\beta$ of two Kramers doublets are
visible. By fixing \vg\ and sweeping a magnetic field, the evolution of the
states in the field can be recorded, see Figs.~\ref{fig:data}(b,c). The Kramers
degeneracy is then lifted, revealing four states in each set.

Low field spectra similar to Fig.~\ref{fig:data}(b) have been reported by 
several groups \cite{minot:nature2004, kuemmeth:nature2008, 
jespersen:natphys2011, steele:natcomms2013} and are now well understood. A 
quantitative fit can be obtained by a model Ha\-miltonian for a single 
longitudinal mode, including valley mixing due to disorder or backscattering at 
the contact (see \cite{jespersen:natphys2011} and Sec.~VI of the Supplement).
For $\left| \Bpar \right| > 0.5\,\text{T}$, valley mixing is
not relevant and the evolution of the spectral lines can be deduced from 
the Dirac equation, Eq.~(\ref{eq:Hgraphene}) below (see Sec.~III of the
Supplement for needed modifications). Valley and spin can
be assigned to each excitation at higher fields, see Fig.~\ref{fig:data}(b).
 
We have traced the single particle states from Fig.~\ref{fig:data}(b) up to a 
high magnetic field of $\Bpar = 17\un{T}$. As visible in
Figs.~\ref{fig:data}(b) and \ref{fig:data}(c), the four $K$ lines evolve upwards
in energy. They are comparatively weak, fading out already below $1\un{T}$. In
contrast, the four $K'$ conductance lines evolve initially downwards, gaining
in strength, but then turn upwards above $6\un{T}$ and fade too. The presence of
both weak $K$ and strong $K'$ transitions in Fig.~\ref{fig:data}(c) at the same
bias excludes the possibility of a trivial dependence of tunneling rates on the
bias voltage. The model calculation of the conductance in
Fig.~\ref{fig:data}(d), assuming a field independent \kpar, successfully
follows the peak positions but clearly fails to reproduce the intensity
variations, especially the suppression of $K$ lines already at low fields.

We show in the following that this effect results from the \Bpar\ dependence of
the wave functions' longitudinal profile. When the field is applied
perpendicular to the CNT axis no such effect occurs and all excitation lines
are present at almost constant strength; see Fig.~S-10 in the Supplement, where
this is experimentally reproduced over a wide gate voltage and electron number
range~\cite{supplement}.

{\em{Boundary conditions on bipartite lattices---}}
The spatial profile of the wave functions $\psi(\ve{r})$ of a finite quantum 
system is determined by the boundary conditions and the resulting quantization 
of the wave vector. In unipartite lattices, e.g., monoatomic chains, hard-wall 
boundary conditions are $\psi(\ve{R}_L) = 0 = \psi(\ve{R}_R)$, where 
$\ve{R}_{L/R}$ are the lattice vectors of the first site beyond the left and 
right end of the chain, respectively. The linear combinations of Bloch states 
satisfying these conditions create standing waves with nodes at $\ve{R}_L$ 
and $\ve{R}_R$, as those of a {\em half-wave} resonator. Their wave vectors are 
quantized according to the familiar condition $k_\parallel = n\pi/L$, where $L$ 
is the length of the chain and $n \in\mathbb{N}$.

\begin{figure}[tp]
\includegraphics[width=\columnwidth]{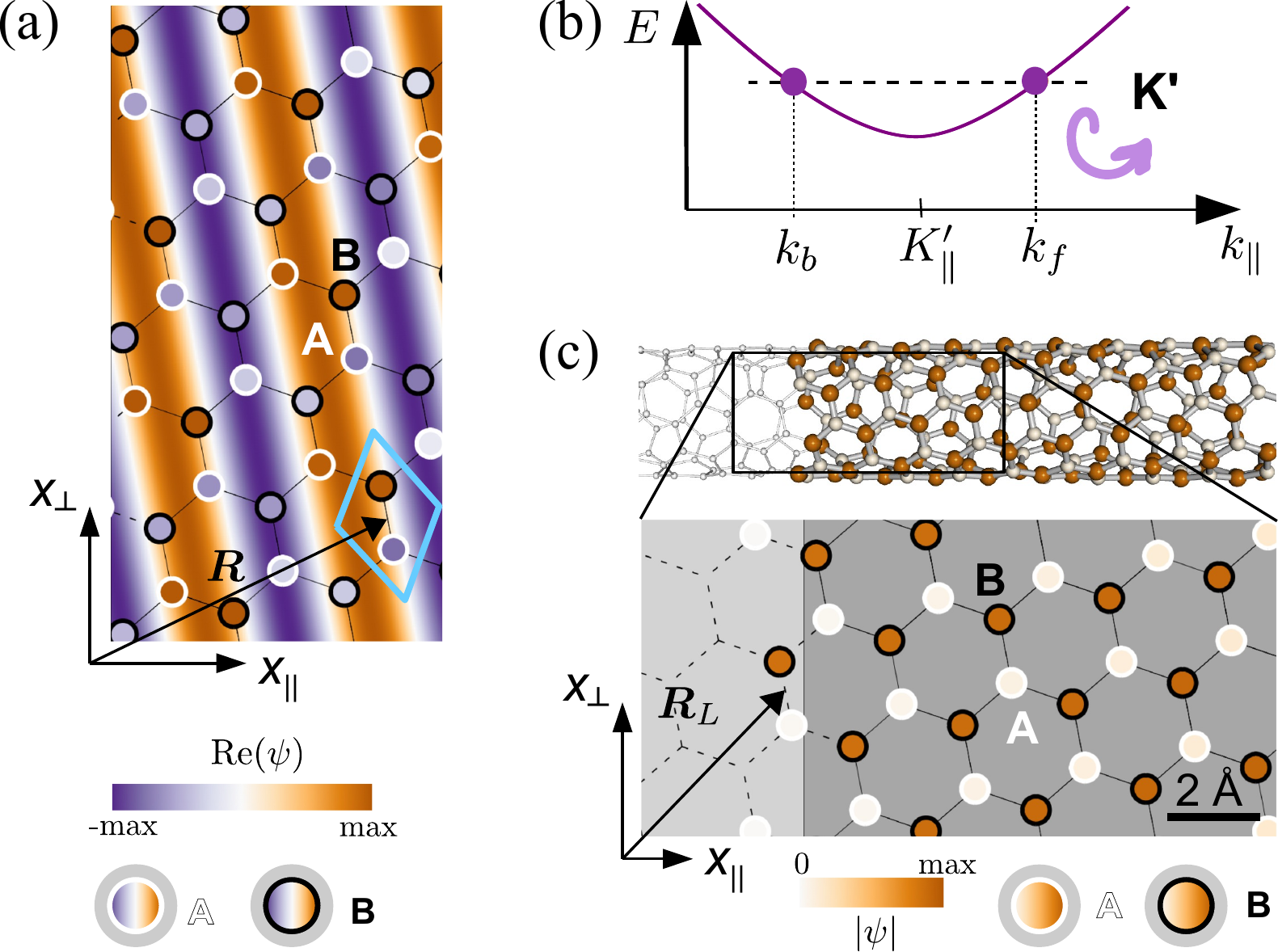}
\caption{\label{fig:boundary}
(a) Bloch function $\psi_{\boldsymbol{k}_{\! f}}$ in a (6,3) infinite CNT 
calculated at the atomic positions (filling of white-rimmed circles for the
A sublattice, black-rimmed for B). The background shows a forward-propagating
[$f$] plane wave with momentum $\boldsymbol{k}_{\! f}$.
(b) 
Level diagram corresponding to forward [$f$] and backward [$b$] propagating 
states in the K' valley. Standing waves in a finite CNT are composed of such 
states from the opposite sides of a Dirac cone at the same energy.
(c) Left end of a (6,3) chiral CNT. The solid-drawn atoms and bonds belong 
to the quantum dot, the faint ones to the tunneling region. The calculated 
amplitude of the energy eigenstate formed by the superposition of 
$\psi_{\boldsymbol{k}_{\! f}}$ and $\psi_{\boldsymbol{k}_b}$ (circle filling / 
atom coloring) approaches zero towards the left end on the A atoms only.
}
\end{figure}
The situation is more complex in bipartite lattices, as in the CNT shown in 
Fig.~\ref{fig:boundary}. The eigenstates are spinors in sublattice space, 
$\Psi^\dag=(\psi_A^\dag,\psi_B^\dag)$, and near the Dirac points obey 
the Dirac equation
\begin{equation}
\label{eq:Hgraphene}
\hbar v_F \!\begin{pmatrix}
        0 \quad\quad e^{i\tau\theta}(\tau\kappa_\perp - i\kappa_\parallel)
        \\[2mm]
        e^{-i\tau\theta}(\tau\kappa_\perp + i\kappa_\parallel) \quad\quad 0
       \end{pmatrix}\!\!\!
       \begin{pmatrix}
       \psi_{\ve{k}A} \\ \psi_{\ve{k}B}
       \end{pmatrix}
= E\!
       \begin{pmatrix}
       \psi_{\ve{k}A} \\ \psi_{\ve{k}B}
       \end{pmatrix}\!,
\end{equation}
where $v_F$ is the Fermi velocity and $\theta$ is the CNT chiral angle. They 
have the form $\Psi_{\ve{k}} = w(e^{i\eta(\ve{k})} 
\psi_{\boldsymbol{k}A}+ e^{-i\eta(\ve{k})} \psi_{\boldsymbol{k}B})$, with $w$ a 
normalization factor, meaning that there is a phase shift $2\eta(\ve{k}) = 
-\tau\arctan(\kappa_\parallel/\kappa_\perp) + \tau\theta$ between the two 
sublattice wavefunctions $\psi_{\ve{k}A}$ and $\psi_{\ve{k}B}$. On the $A$
atoms the phase is advanced by $\eta(\ve{k})$ with respect to the plane wave
part of the Bloch state, on the $B$ atoms it is retarded. This is illustrated in
Fig.~\ref{fig:boundary}(a), where the real part of
the plane wave $e^{i\boldsymbol{k}_{\! f} \cdot \boldsymbol{r}}$ is plotted in 
the background, and the real part of the complete Bloch function 
$\Psi_{\boldsymbol{k}_{\! f}}(\boldsymbol{r})$ at each atomic position is shown 
as the filling of the white (sublattice $A$) and black (sublattice $B$) circles.

Standing waves in a finite CNT are formed by appropriate linear combinations of 
forward [$f$] and backward [$b$] propagating waves of the same energy, see
Fig.~\ref{fig:boundary}(b). A specific combination of Bloch states $\Psi = c_f 
\Psi_{\ve{k}_f} + c_b \Psi_{\ve{k}_b}$ may satisfy the boundary condition 
$\psi_A(\ve{R}_L) = 0$, but then in general $\psi_B(\ve{R}_L)\neq 0$. The 
counterpropagating Bloch waves interfering destructively on $A$ remain finite 
on $B$ because they are superposed with different phases, see
Fig.~\ref{fig:boundary}(c). There is no non-trivial superposition with nodes at 
both ends for both sublattice components. Thus, the boundary conditions for 
bipartite lattices are either $\psi_A(\ve{R}_{L}) = 0 = \psi_B(\ve{R}_{R})$ or 
$\psi_A(\ve{R}_{R}) = 0 = \psi_B(\ve{R}_{L})$, depending on the sublattice to 
which the majority of the relevant edge atoms belongs \cite{akhmerov:prb2008, 
marganska:prb2011, note:armchair}.
\begin{figure}[b]
\includegraphics[width=0.9\columnwidth]{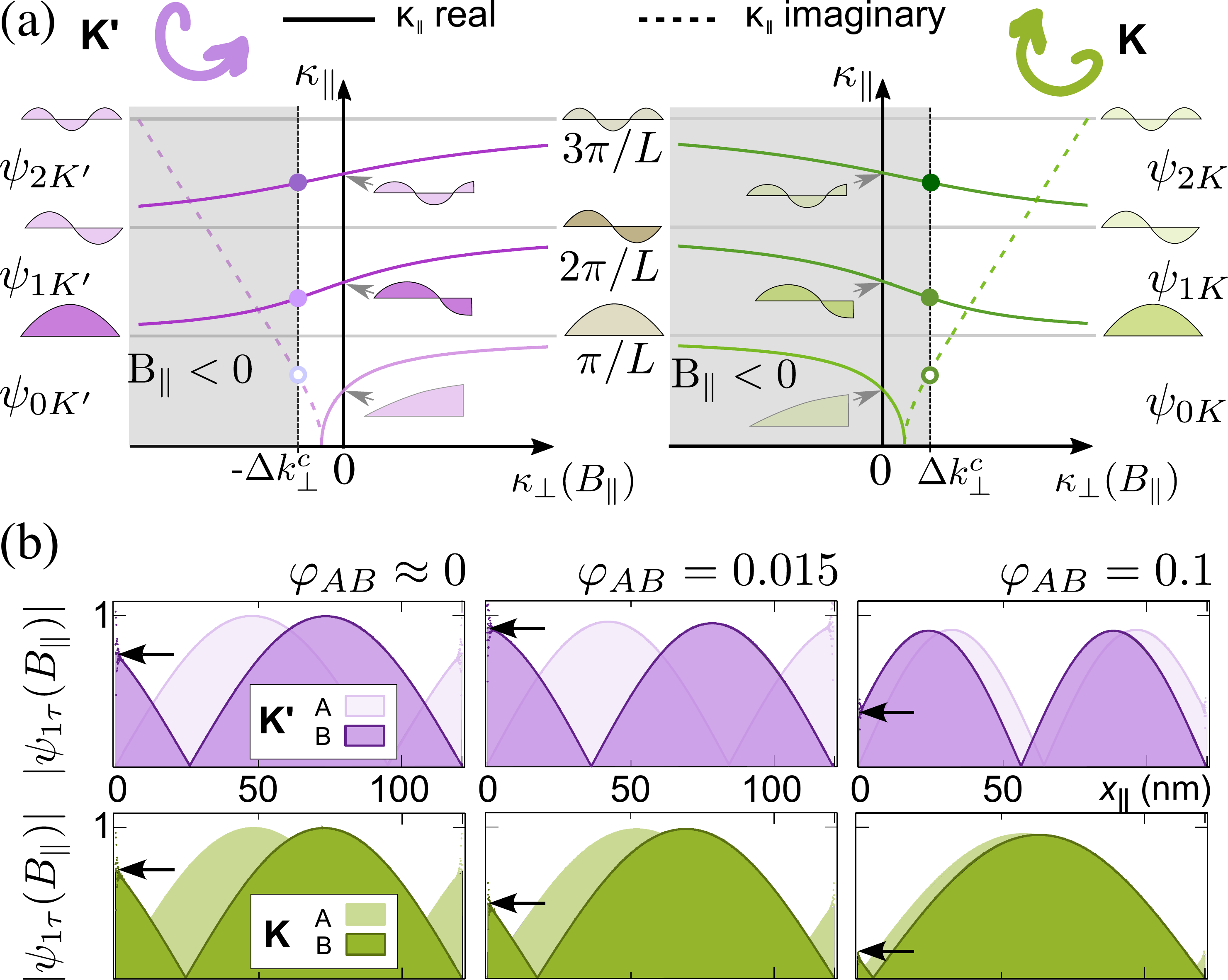}
\caption{\label{fig:theory}
(a) The first solutions of Eq.~(\ref{eq:cross-quantization}) in the $K'$ and 
$K$ valley for a chiral CNT. The wave function envelope is sketched for 
$\kappa_\perp=0$ and large $\kappa_\perp$. Dots mark the values of
$\kappa_\parallel$ at $\Bpar=0$.
(b) Calculated amplitude of the longitudinal wave functions $\psi_{1\tau} 
(x_\parallel, \Bpar)$ of the $K'$ (top row) and $K$ (bottom row) valley states 
for a (15,3) chiral CNT with $L = 121\un{nm}$. Black arrows mark $\vert
\psi_{1\tau} \vert$ on the $B$ sublattice at the left end.
} 
\end{figure}
The superposition of forward and backward moving Bloch states with $\pm 
\kappa_\parallel$ and the same $\tau\kappa_\perp$, together with the bipartite 
boundary conditions, leads to the unusual quantization condition 
\cite{akhmerov:prb2008, castroneto:rmp2009, marganska:prb2011}
\begin{equation}
\label{eq:cross-quantization}
e^{2i\kappa_\parallel L} \overset{!}{=}
e^{-2i\eta(\boldsymbol{k})}\,e^{i\tau\theta} = \frac{\tau \kappa_\perp +
i\kappa_\parallel}{\tau \kappa_\perp - i\kappa_\parallel}.
\end{equation}
Since Eq.~(\ref{eq:cross-quantization}) couples the transverse and the 
longitudinal direction, it can be seen as a {\em cross-quantization} condition. 
It implies that in an axial field also \kpar\ depends on \Bpar. 

The solutions of Eq.~(\ref{eq:cross-quantization}) are plotted as coloured 
lines in Fig.~\ref{fig:theory}(a). For comparison, the grey lines parallel to 
the \kperp\ axis correspond to the familiar half-wave solutions. The envelope
wave function on the $A$ sublattice is also sketched; the $B$ counterpart is
its mirror image. When \kpar\ is close to a multiple of $\pi/L$ (for large
\Bpar), the wave function has the standard half-wave shape with a node at each
end. At low field, the profile on each sublattice is close to a quarter-wave,
with an anti\-node at the corresponding unconstrained end.

Figure~\ref{fig:theory}(b) shows the calculated wave function amplitudes for 
the lowest mode ($n=1$), $|\psi_{1\tau}(x_\parallel)|$ on the $A$ and $B$ 
sublattices, of a (15,3) CNT with $L = 121\un{nm}$. They were obtained by 
direct diagonalization of a tight-binding Hamiltonian on finite lattice, with 
four valence orbitals per atom (for clarity without spin dependence) 
\cite{izumida:jpsj2009, klinovaja:prb2011}. The shapes follow closely the 
expectations based on our analysis of Eq.~(\ref{eq:cross-quantization}). 

\begin{figure}[b]
\includegraphics[width=\columnwidth]{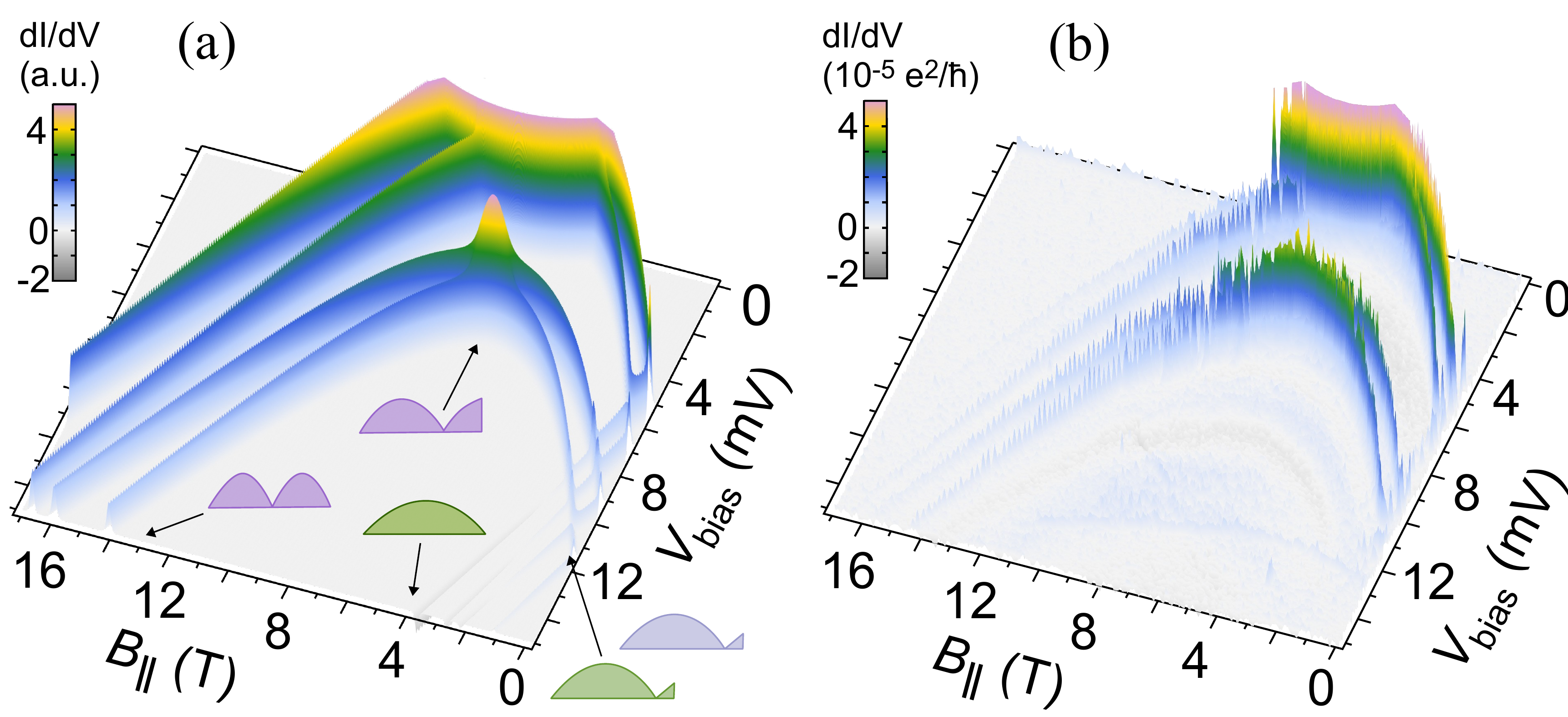}
\caption{\label{fig:exptheo}
(a) Differential conductance calculated using the cross-quantization condition, 
Eq.~(\ref{eq:cross-quantization}), and thus {\it field-dependent} 
tunneling couplings. The wave functions $\psi_{K}(x_\parallel, B_\parallel)$ 
and $\psi_{K'} (x_\parallel, \Bpar)$ are assumed to represent the mode $n=1$, 
with parameters identical for both sets $\alpha$ and $\beta$ (cf. 
Fig.~\ref{fig:data}) and spin independent. 
(b) 3-dimensional plot of the data of Fig.~\ref{fig:data}(c), showing clearly 
the variation of peak heights with \Bpar.
}
\end{figure}
{\em{Fading of the differential conductance---}}
To explain the fading conductance lines in Figs.~\ref{fig:data}(b,c), we account 
for the \Bpar-dependence of the longitudinal CNT wave function in our transport 
calculations. This implies a \Bpar\ dependent tunneling amplitude, given by the 
overlap between CNT and lead wave functions in the contact region. In the
single electron regime of the experiment, tunneling is weak and the tunneling
amplitude is to a good approximation determined by the value of the CNT
wavefunction at the quantum dot ends. The tunnel coupling at the left ($L$)
contact is then
\begin{equation}
\label{eq:Gammas}
\Gamma_{L\mu}(\Bpar)  = 
\alpha_L\frac{2\pi}{\hbar} \vert\psi_{B\mu}(x_{\parallel}=0, 
\Bpar)\vert^2,
\end{equation}
where $\mu =(n,\tau,\sigma)$ is a collective index accounting for the mode, 
valley, and spin, and $\alpha_L$ contains both the square modulus of the lead
wave function at the contact and the lead density of states. The tunnel
coupling at the right ($R$) contact is obtained by replacing $A \leftrightarrow
B$ and $L\leftrightarrow R$. The factors $\alpha_l$ ($l=L,R $) encode a
possible contact asymmetry. The differential conductance then follows from a
reduced density matrix approach to lowest order in $\Gamma_{l\mu}$
\cite{koller:prb2010, supplement}. A calculation assuming $\alpha_L / \alpha_R
= 1/4$ is shown in Fig.~\ref{fig:exptheo}(a). The input parameters for Eqs.
(\ref{kappapar}) and (\ref{eq:cross-quantization}) (nanotube radius, length,
and $\Delta k_\perp^c$) were obtained by fitting the measured position of the
spectral lines shown in Fig.~\ref{fig:data}(b,c) to the spectrum of the CNT
model Hamiltonian, see Sec. III of the Supplement. The fast disappearance
of the $K$ lines is in excellent agreement with the data plotted in 
Fig.~\ref{fig:exptheo}(b). The suppression of $K'$ lines at high field is also 
clearly reproduced. 

In our calculations hard wall boundary conditions were assumed. In the
experiment, though, we expect smooth confinement due to electrostatic gating,
cf. Fig.~S-6 of the Supplement. Hence, we have performed numerical calculations
of the CNT eigenmodes as a function of \Bpar\ for a soft confinement, see 
Sec.~V of the Supplement \cite{supplement}. We find qualitative agreement with
the hard wall confinement calculation. Thus, the tunability of the longitudinal
wave function with magnetic field occurs for smooth confinement as well.

In conclusion, our experiment can be regarded as the complement of a scanning 
tunneling microscopy (STM) measurement. In STM the spatial profile of atomic or 
molecular orbitals is obtained by scanning the tip position over the sample. In 
CNT quantum dots, the contact position is fixed, but the wavefunction, and thus 
the tunnel current, is tuned by an axial magnetic field. We are aware of only 
one other system in which such coupling has been found, a semiconducting 
quantum dot with pyramid shape \cite{cao:scirep2015}. The unusual tunability of 
the wave function shape with a parallel magnetic field will influence all 
phenomena dependent on the full spatial profile of the electronic states, such 
as, e.g., electron-phonon coupling or electron-electron interaction. Thus the 
parallel magnetic field is an even more versatile tool to investigate and 
control complex quantum systems than already acknowledged.

\begin{acknowledgments}
The authors thank the Deutsche Forschungsgemeinschaft for financial support via 
SFB 689, SFB 1277, GRK 1570, and Emmy Noether grant Hu 1808/1. We also thank S.
Ilani for stimulating discussions. The measurement data has been recorded using
the \href{https://www.labmeasurement.de/}{Lab::Measurement} software package
\cite{l-m}.
\end{acknowledgments}


\begin{thebibliography}{39}
\bibitem [{not()}]{note:armchair}%
  \BibitemOpen
  \href@noop {} {}\bibinfo {note} {The only exception is an armchair nanotube,
  which has equal numbers of $A$ and $B$ atoms at the edges and obeys the usual
  half-wave quantization condition $k_\parallel = \pi/L$.}\BibitemShut {Stop}%
\end{thebibliography}
\end{document}